%% file: main.tex
\documentclass{article}

\input{mypackages.tex}

\input{mystyle.tex}

\title{Bullshark: The Partially Synchronous Version}

\author[1]{Alexander Spiegelman}
\author[2]{Neil Giridharan}
\author[3]{Alberto Sonnino}
\author[4]{Lefteris Kokoris-Kogias}
\affil[1]{Aptos}
\affil[2]{University of California, Berkeley}
\affil[3]{Mysten Labs}
\affil[4]{IST Austria}

\date{}

\begin{document}

\maketitle

\begin{abstract}

The purpose of this manuscript is to describe the deterministic partially synchronous version of Bullshark in a simple and clean way. \textbf{This result is published in CCS 2022}, however, the description there is less clear because it uses the terminology of the full asynchronous Bullshark. The CCS version ties the description of the asynchronous and partially synchronous versions of Bullshark since it targets an academic audience. Due to the recent interest in DAG-based BFT protocols, we provide a separate and simple description of the partially synchronous version that targets a more general audience. 
We focus here on the DAG ordering logic. For more details about the asynchronous version, garbage collection, fairness, proofs, related work, evaluation, and efficient DAG implementation please refer to~\cite{fullpaper, ccs, danezis2021narwhal, keidar2021all}.
For an intuitive extended summary please refer to the blogpost~\cite{DAGmeetsBFT}.

\end{abstract}

\input{intro}

\input{protocol}

\clearpage

\bibliographystyle{plain}
\bibliography{bib}

\end{document}

%% file: mypackages.tex
\usepackage{amsmath}
\usepackage{amssymb}
\usepackage{amsthm}
\usepackage{graphicx}
\usepackage[colorinlistoftodos]{todonotes}
\usepackage{algorithm}
\usepackage[noend]{algpseudocode}
\usepackage{enumerate}
\usepackage{xcolor}
\usepackage{framed}
\usepackage{bm}
\usepackage{pifont}
\usepackage{multicol}
\usepackage{lipsum}
\usepackage{tikz}
\usetikzlibrary{calc}
\usepackage{wrapfig}
\usepackage{array}
\usepackage{relsize}
\usepackage{comment}
\usepackage{url}
\usepackage[title]{appendix}
\usepackage{caption}
\usepackage{subcaption}
\usepackage{xstring}
\usepackage{thmtools,thm-restate}
\usepackage{fullpage}
\usepackage{authblk}

\usepackage{nicefrac}
\usepackage{siunitx}
\usepackage{amsmath}
\usepackage{amsfonts}
\usepackage{array,framed}
\usepackage{booktabs}
\usepackage{multicol}

\usepackage{
  color,
  float,
  epsfig,
  wrapfig,
  graphics,
  graphicx,
  subcaption
}
 \usepackage{setspace}
 \usepackage{latexsym,fancyhdr,url}
\usepackage{enumerate}
\usepackage{algorithm}
\usepackage[noend]{algpseudocode}
\usepackage{graphics}
\usepackage{xparse} 
\usepackage{xspace}
\usepackage{multirow}
\usepackage{csvsimple}
\usepackage{balance}
\usepackage{lineno}

\usepackage{
  tikz,
  pgfplots,
  pgfplotstable
}
\usepackage{hyperref}

\usetikzlibrary{
  shapes.geometric,
  arrows,
  external,
  pgfplots.groupplots,
  matrix
}

\usepackage{booktabs}       

%% file: mystyle.tex
\algdef{SE}[RECEIVING]{Receiving}{EndReceiving}[1]{\textbf{upon
receiving}\ #1\ \algorithmicdo}{\algorithmicend\ \textbf{}}%
\algtext*{EndReceiving}

\algdef{SE}[UPON]{Upon}{EndUpon}[1]{\textbf{upon
}\ #1\ \algorithmicdo}{\algorithmicend\ \textbf{}}%
\algtext*{EndUpon}

\algdef{SE}[DELIVERY]{Delivery}{EndDelivery}[1]{\textbf{upon
delivery}\ #1\ \algorithmicdo}{\algorithmicend\ \textbf{}}%
\algtext*{EndDelivery}

\usepackage{mathtools,}

\newcommand\StateX{\Statex\hspace{\algorithmicindent}}
\newcommand\StateXX{\StateX\hspace{\algorithmicindent}}
\algrenewcommand\textproc{}

\newcommand{\alglinenoNew}[1]{\newcounter{ALG@line@#1}}
\newcommand{\alglinenoPop}[1]{\setcounter{ALG@line}{\value{ALG@line@#1}}}
\newcommand{\alglinenoPush}[1]{\setcounter{ALG@line@#1}{\value{ALG@line}}}

\newcommand{\sys}{BullShark\xspace}

\DeclareMathAlphabet{\mathcal}{OMS}{cmsy}{m}{n}
\DeclareGraphicsExtensions{%
    .png,.PNG,%
    .pdf,.PDF,%
    .jpg,.mps,.jpeg,.jbig2,.jb2,.JPG,.JPEG,.JBIG2,.JB2}

\algdef{SE}[Upon]{Upon}{EndUpon}[1]{\textbf{upon}\ #1\ \algorithmicdo}{\algorithmicend\ \textbf{}}%
\algtext*{EndUpon}

\usepackage{xparse}
\newcommand{\bnm}{\begin{newmath}}
\newcommand{\enm}{\end{newmath}}

\newcommand{\bea}{\begin{eqnarray*}}%
\newcommand{\eea}{\end{eqnarray*}}%

\newcommand{\bne}{\begin{newequation}}
\newcommand{\ene}{\end{newequation}}

\newcommand{\bal}{\begin{newalign}}
\newcommand{\eal}{\end{newalign}}

\newenvironment{newalign}{\begin{align}%
\setlength{\abovedisplayskip}{4pt}%
\setlength{\belowdisplayskip}{4pt}%
\setlength{\abovedisplayshortskip}{6pt}%
\setlength{\belowdisplayshortskip}{6pt} }{\end{align}}

\newenvironment{newmath}{\begin{displaymath}%
\setlength{\abovedisplayskip}{4pt}%
\setlength{\belowdisplayskip}{4pt}%
\setlength{\abovedisplayshortskip}{6pt}%
\setlength{\belowdisplayshortskip}{6pt} }{\end{displaymath}}

\newenvironment{newequation}{\begin{equation}%
\setlength{\abovedisplayskip}{4pt}%
\setlength{\belowdisplayskip}{4pt}%
\setlength{\abovedisplayshortskip}{6pt}%
\setlength{\belowdisplayshortskip}{6pt} }{\end{equation}}

\newcounter{ctr}

\newcounter{mytable}
\def\mytable{\begin{centering}\refstepcounter{mytable}}
\def\endmytable{\end{centering}}

\newcounter{myfig}
\def\myfig{\begin{centering}\refstepcounter{myfig}}
\def\endmyfig{\end{centering}}

\newlength{\saveparindent}
\newlength{\saveparskip}
\newcommand{\E}{{\rm I\kern-.3em E}}

\renewcommand{\eqref}[1]{\mbox{Equation~(\ref{#1})}}

\def \part {part}

\renewcommand{\paragraph}[1]{\vspace*{6pt}\noindent\textbf{#1}\;}

\def \blackslug{\hbox{\hskip 1pt \vrule width 4pt height 8pt
    depth 1.5pt \hskip 1pt}}
\def \qed{\quad\blackslug\lower 8.5pt\null\par}

\newcounter{mynote}[section]

\newcommand\ignore[1]{}

\newcounter{rcnote}[section]

\newcounter{mrnote}[section]

\newcounter{fknote}[section]

\newcounter{anote}[section]

\DeclareMathSymbol{\mlq}{\mathord}{operators}{``}
\DeclareMathSymbol{\mrq}{\mathord}{operators}{`'}

\newcommand{\rhf}[2]{R_{f, \gamma}}

\DeclareDocumentCommand{\edist}{o o}{
  \ensuremath{
    \IfNoValueTF{#1}{{d}}{{\sf d}(#1,#2)}
  }
}

\newcommand{\olrk}[1]{\ifx\nursymbol#1\else\!\!\mskip4.5mu plus 0.5mu\left(\mskip0.5mu plus0.5mu #1\mskip1.5mu plus0.5mu \right)\fi}

\NewDocumentCommand{\indseq}{ O{1} O{r} }{{#1}\ldots {#2}}


%% file: intro.tex
\section{Introduction}

In the context of Blockchains, BFT consensus is a problem in which $n$ parties, $f$ of which might be Byzantine, try to agree on an infinitely growing sequence of transactions. The idea of DAG-based BFT consensus is to separate the network communication layer from the ordering (consensus) logic. Each message contains a set of transactions, and a set of references to previous messages. Together, all the messages form a DAG that keeps growing – a message is a vertex and its references are edges.
The networking layer of building the DAG can and should be optimized on a system level (see Narwhal~\cite{danezis2021narwhal}).
Once a DAG is constructed the ordering logic of its vertices adds zero communication overhead. That is, each party independently looks at its local view of the DAG and totally (fully) orders all the vertices without sending a single extra message. This is done by interpreting the structure of the DAG as a consensus protocol, i.e., a vertex can be a proposal and an edge can be a vote.
Importantly, due to the asynchronous nature of the network, different parties may see slightly different DAGs at any point in time. A key challenge then is how to guarantee that all the parties agree on the same total order.

%% file: protocol.tex
\section{Protocol}

We describe here the partially synchronous version of Bullshark. 
To focus on the ordering logic of the DAG (Section~\ref{sub:order}), we first assume a DAG with certain properties~\ref{sub:prop} is given.
Then, for completeness, we discuss a few alternatives to construct the DAG (Section~\ref{sub:construction}).
Details about the fairness, garbage collection, evaluation can be found in~\cite{fullpaper, ccs, DAGmeetsBFT}, and an efficient DAG implementation in~\cite{danezis2021narwhal}.
Check~\cite{fullpaper} for rigorous (safety and liveness) proofs.

\subsection{DAG properties.}
\label{sub:prop}
\begin{figure}
    \centering
    \includegraphics[width=0.7\textwidth]{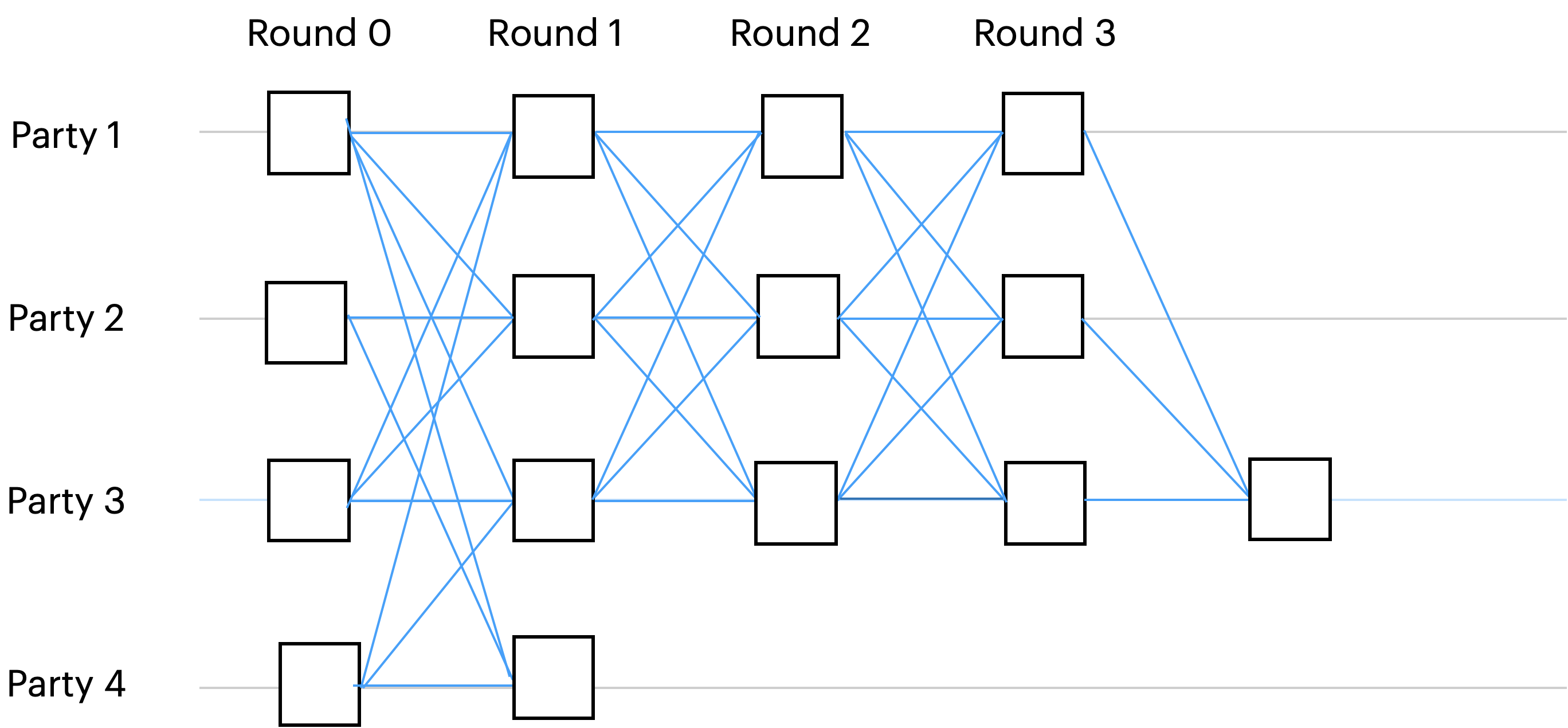}
    \caption{Illustration of the DAG structure.}
    
    \label{fig:DAG}
\end{figure}

We consider a round-based DAG, see illustration in Figure~\ref{fig:DAG}. 
Each round contains at most $n$ vertices (at most 1 vertex per party). Each vertex is associated with a \emph{round} number and its \emph{source} (the party that broadcast it). In addition to transactions information, each vertex also contains a set of at least $n-f$ \emph{edges}, which point to vertices in the previous round.

We discuss the construction of the DAG below, but it is important to note now that due to the asynchronous nature of the network, different parties may see slightly different views of the DAG at any point in time. 
In order to guarantee that all the parties agree on the same total order of the DAG's vertices without any extra communication, the construction of the DAG must satisfy the following:

\begin{description}

\item \textbf{Validity:} if an honest party has a vertex $\mathsf{v}$ in its local view of the DAG, then it also has all the causal history of $\mathsf{v}$ (all vertices that can be reached from $\mathsf{v}$).

\item \textbf{Reliability:} if an honest party has a vertex in round $\mathsf{r}$ by party $\mathsf{p}$ in its local view of the DAG, then eventually all honest parties has a vertex in round $\mathsf{r}$ by party $\mathsf{p}$ in their views of the DAG. 

\item \textbf{Non-equivocation:} if two honest parties have a vertex in round $\mathsf{r}$ by party $\mathsf{p}$ in their local views of the DAG, then the vertices are identical (i.e., transaction information and edges are exactly the same).

\end{description}

\noindent By recursively applying Validity and Non-equivocation we get:

\begin{description}

\item \textbf{Completeness:} if two honest parties have a vertex $\mathsf{v}$ in round $\mathsf{r}$ by party $\mathsf{p}$ in their local views of the DAG, then $\mathsf{v}$'s causal histories are identical in both parties' local view of the DAG.

\end{description}

\subsection{Ordering logic.}
\label{sub:order}
The non-equivocation property of the DAG eliminates the ability of Byzantine parties to lie, and as a result, the completeness property allows us to apply deterministic logic to order the DAG even though parties have slightly different local views of the DAG, which drastically simplifies the ordering logic. 

As mentioned above each party observes its local view of the DAG and totally orders its vertices without any extra communication.
\textbf{Notably, Bullshark needs neither a view-change nor a view-synchronization mechanism.}
Since the DAG encodes full information, there is no need to “agree” on skipping and discarding slow/faulty leaders via timeout complaints and view-synchronization we get for free by the nature of the DAG construction.

We next demonstrate how each party locally interprets the structure of its view of the DAG via a running example with $n=4$ and $f=1$. A detailed pseudocode appears in Algorithms~\ref{alg:dataStructures}.
and~\ref{alg:ESBullshark} (the function \textsc{TryCommitting}($\mathsf{v}$) is invoked by party $p_i$ when a new vertex in an even round is added to its local view of the DAG). 

\begin{figure}[h]
    \centering
    \includegraphics[width=0.6\textwidth]{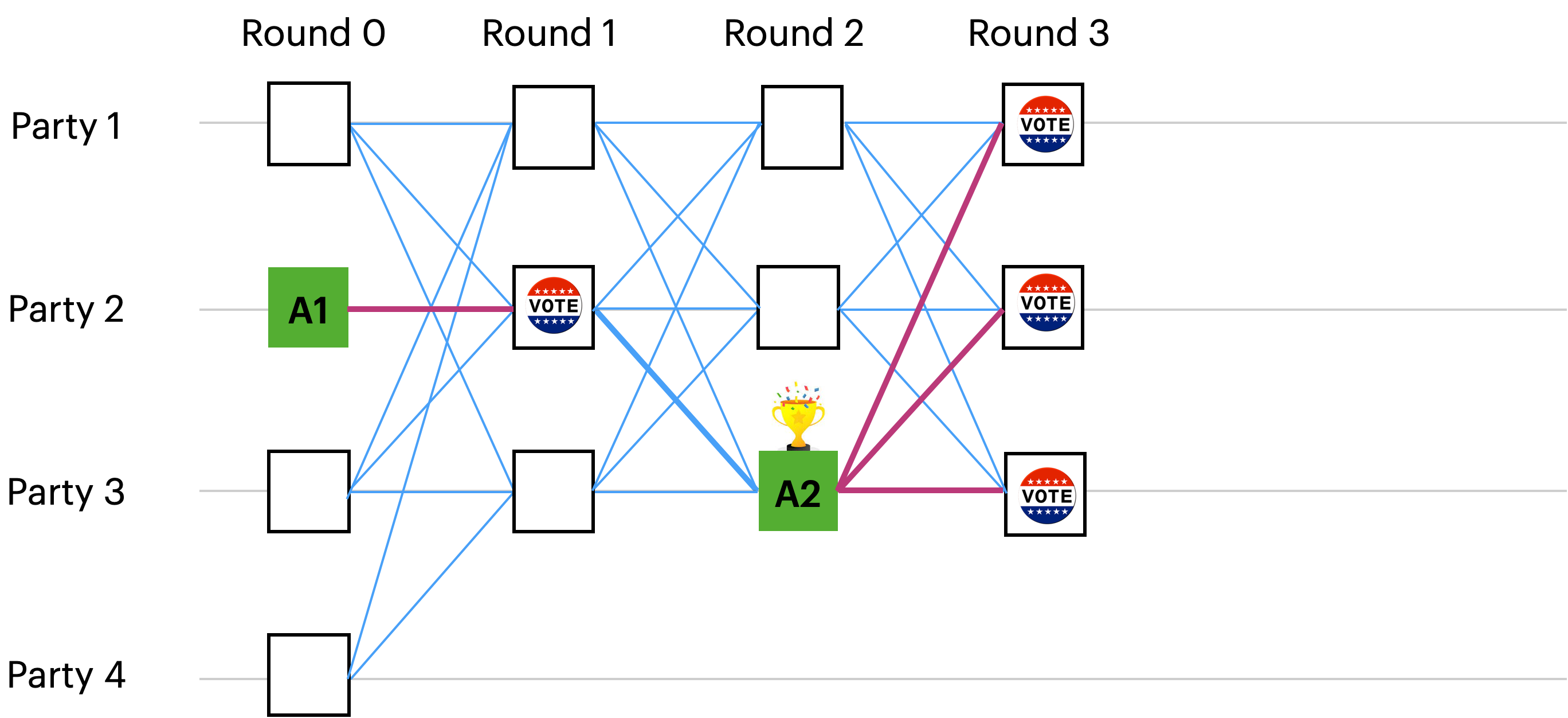}
    \caption{Commit rule requires $f+1$ votes. }
    
    \label{fig:commit}
\end{figure}

Every even-numbered round in the DAG has a predefined leader and we refer to the vertex associated with the leader as anchor. In figure~\ref{fig:commit}, anchors are highlighted in solid green.
The goal is to first decide which anchors to commit. Then, to totally order all the vertices in the DAG, a party goes one by one over all the committed anchors and orders their causal histories by some deterministic rule. Anchor $A2$ causal history is marked by green-outlined vertices.

\begin{figure}
    \centering
    \includegraphics[width=0.6\textwidth]{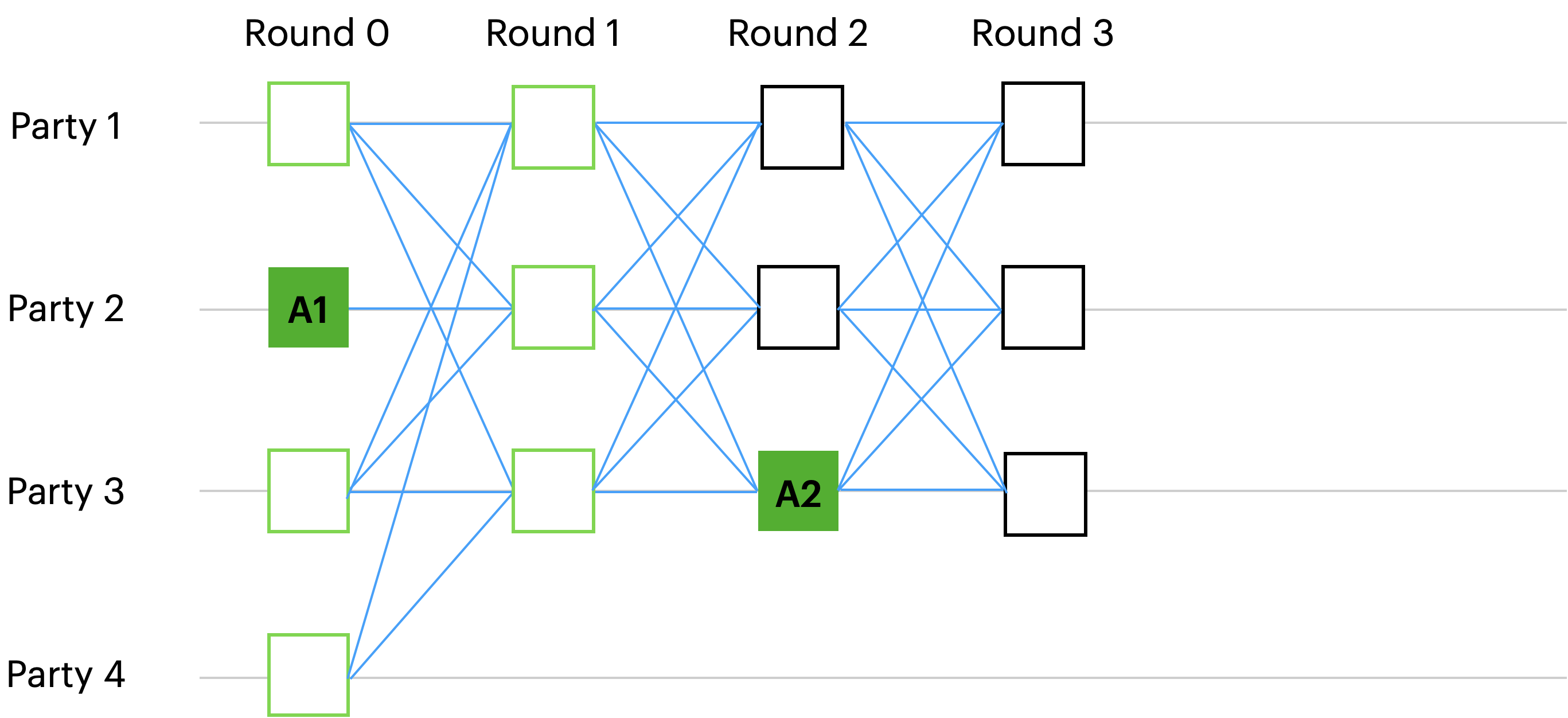}
    \caption{Anchors and causal history.}
    
    \label{fig:anchor}
\end{figure}

Each vertex in an odd round can contribute one vote for the anchor in previous round. 
In particular, a vertex in round $\mathsf{r}$ votes for the anchor in round $\mathsf{r}-1$ if there is an edge between them. 
\textbf{The commit rule is simple:} an anchor is \emph{committed} if it has at least $f+1$ votes. In Figure~\ref{fig:commit}, anchor $A2$ is committed with $3$ votes, whereas anchor $A1$ only has 1 vote and is not committed.

Recall that due to the asynchronous nature of the network, the local views of the DAG might differ for different parties. That is, some vertices might be added to the local view of the DAG of some of the parties but not yet added by the others. 
Therefore, even though some parties have not committed A1, others might have.
In Figure~\ref{fig:localview}, party $p_2$ sees  $2 = f+1$ votes for anchor $A1$ and thus commits it even though party $p_1$ has not. Therefore, to guarantee total order (safety), party $p_1$ has to order anchor $A1$ before anchor $A2$. To achieve this, Bullshark relies on quorum intersection:

\begin{figure}
    \centering
    \includegraphics[width=0.7\textwidth]{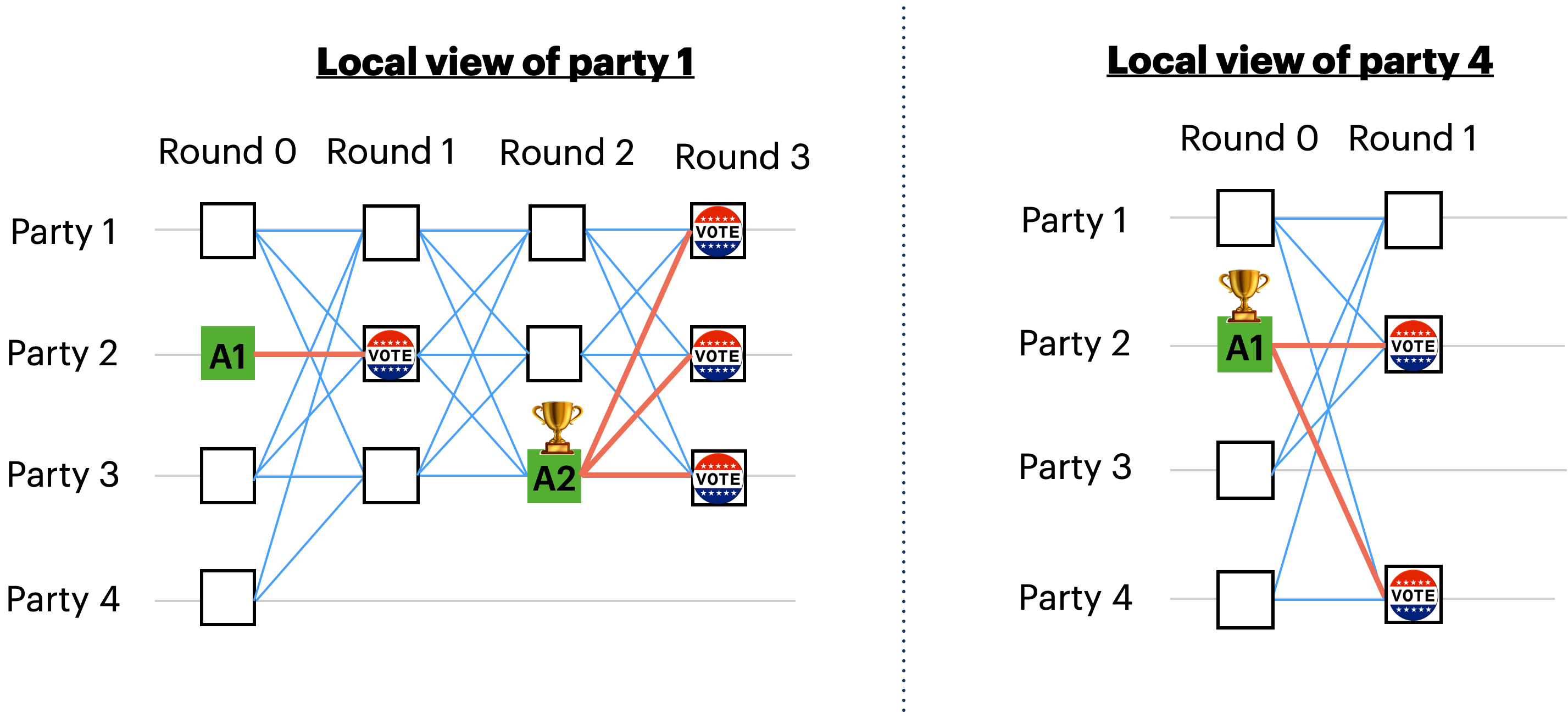}
    \caption{Parties might have slightly different views of the DAG.}
    
    \label{fig:localview}
\end{figure}

\begin{quote}
\textit{Since the commit rule requires $f+1$ votes and each vertex in the DAG has at least $n-f$ edges to vertices from the previous round, it is guaranteed that if some party commits an anchor $A$ then all anchors in higher rounds will have a path to at least one vertex that voted for $A$, and thus will have a path to $A$.}
\end{quote}

\noindent Therefore, we get the following corollary:

\begin{quote}
 \textit{\textbf{If there is no path to a anchor $A$ from a future anchor, then no party committed $A$ and it is safe to skip it.}}
\end{quote}

The mechanism to order anchors is the following: when an anchor $A_i$ is committed, the validator checks if there is a path between anchor $A_i$ to $A_{i-1}$. If this is the case, anchor $A_{i-1}$ is ordered before $A_{i}$ and the mechanism is recursively restarted from $A_{i-1}$. Otherwise, anchor $A_{i-1}$ is skipped and the validator checks if there is a path between $A_{i}$ to $A_{i-2}$. If there is a path, $A_{i-2}$ is ordered before $A_{i}$ and the mechanism is recursively restarted from $A_{i-2}$. Otherwise, anchor $A_{i-2}$ is skipped and the process continues in the same way. The process stops when it reaches an anchor that was previously ordered (as all the anchors before it are already ordered).

\begin{figure}[h]
    \centering
    \includegraphics[width=0.6\textwidth]{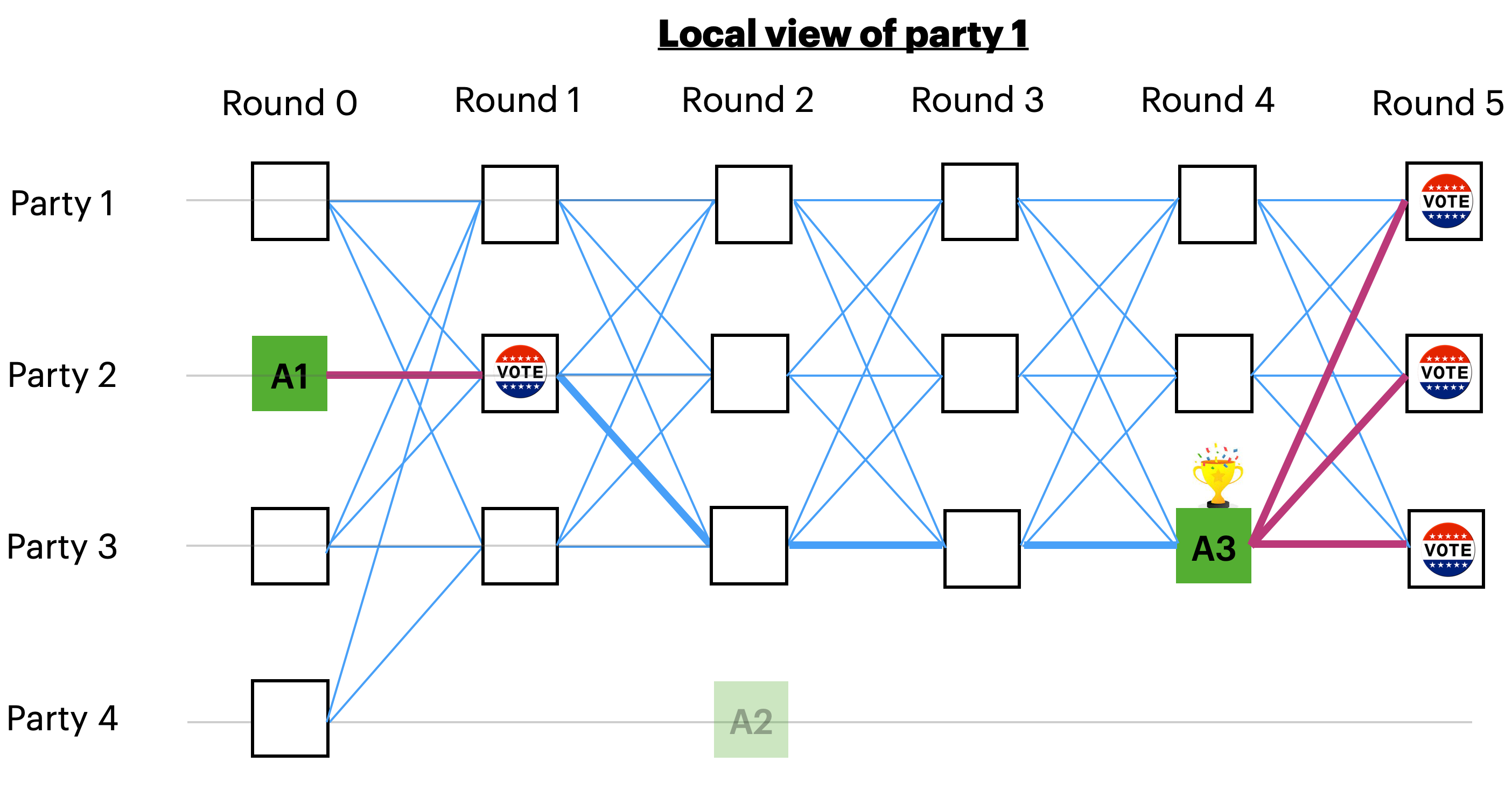}
    \caption{Anchors' ordering logic. $A2$ is not part of party $1$ local view if the DAG}
    
    \label{fig:ordering}
\end{figure}

In Figure~\ref{fig:ordering}, anchors $A1$ and $A2$ do not have enough votes to be committed and once the party commits $A3$ it has to decide whether to order $A1$ and $A2$ before $A3$. Since there is no path from $A3$ to $A2$, $A2$ can be skipped (no party committed it). However, since there is a path between $A3$ and $A1$, $A1$ is ordered before $A3$ ). In this example the process stops here because $A1$ is the first anchor, but in general the process should continue recursively from $A1$ until a previously ordered anchor is reached. 
Note that it might be the case that no party committed $A1$, however it is safe to order it before $A3$ due to the Completeness property of the DAG. All parties see the same history of $A3$ in their local vies of the DAG and thus deterministically agree to order $A1$ before $A3$. In~\cite{fullpaper} we prove by induction that all parties order the same anchors. 

Finally, to totally order the vertices of the DAG, the party orders the causal history of the ordered anchors one by one by some deterministic rule. In our example, it first order $A1$ and then orders the causal history of $A3$. For an intuitive explanation of garbage collection and fairness, and for a DAG construction discussion please refer to the blogpost~\cite{DAGmeetsBFT}.

\newpage
\input{data_structres}

\input{code}

\subsection{DAG construction}
\label{sub:construction}

One naive way to disseminate the vertices and satisfy the DAG properties we need for ordering is to use reliable broadcast~\cite{bracha1987asynchronous}. However, standard implementations require $O(n^2)$ communication per vertex.
We use the Narwhal mempool system, which leverages a proof of availability mechanism, for an efficient and scalable implementation. 
The proof of availability mechanism allows Narwhal to separate data dissemination from the DAG construction, both of which is implemented with $O(n)$ communication complexity in the common case.
In brief, parties keep steaming data to each other. A proof of a availability is a metadata (hash and a quorum of signatures) guaranteeing any party is able to retrieve the data. As a result, instead of blocks of transactions, each vertex of the DAG contains a list of proofs. After ordering the DAG, any party can retrieve the data in case it does not have it.

\paragraph{Advancing rounds.}
Each vertex in the DAG points to at least $n-f$ vertices in the previous round. Therefore, before advancing to round $r$ and broadcasting a new vertex, a party must wait for $n-f$ vertices in round $r-1$. 
Since there are at least $n-f$ honest parties, this can be done completely asynchronously allowing the DAG to grow in network speed.
Indeed, this is exactly how asynchronous solutions like Aleph~\cite{gkagol2019aleph}, Tusk~\cite{danezis2021narwhal} and DAG-Rider~\cite{keidar2021all} advance rounds.

However, as we know from the famous FLP~\cite{fischer1985impossibility} result, deterministic partially synchronous consensus protocols has to use timeouts to satisfy liveness.
Specifically, the problem in advancing rounds whenever $n-f$ vertices are delivered is that parties might not vote for the anchor even if the party that broadcast it is just slightly slower than the fastest $n-f$ parties.
To deal with this, Bullshark integrates timeouts into the DAG construction. 
If the first $n-f$ vertices a party $p$ gets in an even-numbered round $r$ do not include the anchor of round $r$, then $p$ sets a timer and waits for the anchor until the timer expires.
Similarly, in an odd-numbered round, parties wait for either $f+1$ vertices that vote for the anchor, or $2f+1$ vertices that do not, or a timeout.

\paragraph{Responsiveness.}
It is important to note that the DAG construction satisfies responsiveness. That is, after GST, if the leader (the party that broadcasts the anchor) is honest, then the DAG is constructed in network speed and timeouts never expire. 
Moreover, since the Narwhal system disseminates data in network speed regardless the DAG construction, slightly slowing down the DAG construction does not effect the overall throughput (each vertex will simple have slightly more metadata). 

\paragraph{Virtual DAG.}
One alternative we consider is to separate a physical DAG from a virtual DAG. In brief, the idea is that the physical DAG keeps advancing in network speed, i.e., parties advance physical rounds immediately after $n-f$ vertices are delivered in the current round. The logical DAG is piggybacked on top of the physical DAG - each vertex has an additional bit that indicates whether the vertex belongs to the logical DAG. A logical edge between two logical vertices exists if there is a physical path between them. The timeouts are integrated into the logical DAG similarly to how they are integrated into the DAG in the description above. The consensus logic runs on top of the logical DAG and once an anchor is committed, its entire causal history on the physical DAG is ordered.

One may expect that the logical DAG approach would increase the system throughput because parties never wait for a timeout to broadcast vertices on the physical DAG. However, as we explained above, since Narwhal separates data dissemination from the DAG construction, this makes no difference in practice. 
On the other hand, the latency in the logical DAG approach may actually increase – this is because if a logical vertex is not ready to be piggybacked when the physical vertex is broadcast, it will need to wait for the next physical round.
Moreover, from the experience we gained implementing DAG-based protocols, we learned that slightly back pressuring (slowing down) the DAG construction leads to maximum performance since the vertices contain more metadata, which amortizes the overhead of building the DAG (e.g., networking, computation, memory, storage, etc).
In fact, during the development we evaluated and compared both approaches and chose the inject timeouts in the physical layer as it significantly outperformed the decoupled solution (logical DAG) in both throughput and latency.  

For more details please refer to the blogpost~\cite{DAGmeetsBFT}.

%% file: data_structres.tex
\begin{algorithm*}[t]
    \caption{Data structures and basic utilities for party $p_i$}
    \label{alg:dataStructures}
    \begin{algorithmic}[1]
    \small

        \Statex \textbf{Local variables:}
        \StateX struct $\textit{vertex } \mathsf{v}$: 
        \Comment{The struct of a vertex in the DAG}
        \StateXX $ \mathsf{v}.\textit{round}$ - the round of $ \mathsf{v}$ in the DAG
        \StateXX $ \mathsf{v}.\textit{source}$ - the party that broadcast $ \mathsf{v}$
        \StateXX $ \mathsf{v}.\textit{block}$ - a block of transactions information
        \StateXX $ \mathsf{v}.\textit{edges}$ - a set of at least $n-f$ vertices in
        $ \mathsf{v}.\textit{round}-1$ 
        \Comment{Used to provide fairness}
        \StateX $DAG_i[]$ - An array of sets of vertices

        \vspace{0.5em}
        \Procedure{\textsc{path}}{$ \mathsf{v}, \mathsf{u}$} \Comment{Check if exists a path from $\mathsf{v}$ to $\mathsf{u}$ in the DAG}
        \State \Return exists a sequence of $ \mathsf{k} \in \mathbb{N}$,
        vertices $ \mathsf{v_1},  \mathsf{v_2},\ldots,  \mathsf{v_k}$  s.t.\
        \StateXX $ \mathsf{v_1} =  \mathsf{v} $, $ \mathsf{v_k} =  \mathsf{u}$, and $\forall j \in [2..k]
        \colon  \mathsf{v_j} \in \bigcup_{\mathsf{r} \geq 1}
        \mathsf{DAG_i[r]} \wedge \mathsf{v_j} \in  \mathsf{v_{j-1}}.\textit{edges} $
        \EndProcedure


        \vspace{0.5em}
        \Procedure{\textsc{getAnchor}}{$\mathsf{r}$}
            \State  $\mathsf{p} \gets \textsc{getPredefinedLeader}(\mathsf{r})$ 
            \Comment{Assume some predefined mapping known to all parties}
            \If{$\exists \mathsf{v} \in \mathsf{DAG[r]}$ s.t.\ $\mathsf{v}.source = \mathsf{p}$}
                \State \Return $\mathsf{v}$ 
            \EndIf 
                \State \Return $\bot$ 
        \EndProcedure

         \alglinenoNew{counter}
         \alglinenoPush{counter}

    \end{algorithmic}
\end{algorithm*}

%% file: code.tex
\begin{algorithm}[H]
\caption{Eventually synchronous \sys: algorithm for party $p_i$.}
\begin{algorithmic}[1]
\alglinenoPop{counter}
\small

        \Statex \textbf{Local variables:}
      
        \StateX $\mathsf{orderedVertices} \gets \{\}$
        \StateX $\mathsf{lastOrderedRound} \gets 0$
        \StateX $\mathsf{orderedAnchorsStack} \gets $ initialize empty stack
    
       \Statex

        \Procedure{\textsc{TryCommitting}}{$\mathsf{v}$}
         \If{$\mathsf{v}.\textit{round} \text{ mod }2 = 1$  or $\mathsf{v}.\textit{round} = 0$}
            \State return 
        \EndIf    
    
        \State $\mathsf{anchor} \gets$ \textsc{getAnchor($\mathsf{v}.\textit{round-2}$)}
        \State $\mathsf{votes} \gets \mathsf{v}.\textit{edges}$
        \If{$|\{ \mathsf{v'} \in \mathsf{votes}: \textsc{path}(\mathsf{v'},\mathsf{anchor})\}|
        \geq f+1$}
        
        \State \textsc{orderAnchors($\mathsf{anchor}$)}
        
        \EndIf

    \EndProcedure
       
     \StateX

    \Procedure{\textsc{orderAnchors}}{$\mathsf{v}$}
        \State $\mathsf{anchor} \gets \mathsf{v}$
        \State $\mathsf{orderedAnchorsStack}.\textit{push}(\mathsf{anchor})$
        \State $\mathsf{r} \gets \mathsf{anchor}.\textit{round} -2$
        \While{$\mathsf{r} > \mathsf{lastOrderedRound}$}
        \State $\mathsf{prevAnchor} \gets \textsc{getAnchor}(\mathsf{r})$
            
            \If{$\textsc{path}(\mathsf{anchor}, \mathsf{prevAnchor})$}
            \State $\mathsf{orderedAnchorsStack}.\textit{push}(\mathsf{prevAnchor})$
            \State $\mathsf{anchor} \gets \mathsf{prevAnchor}$ 
            
        \EndIf
        
        \State $\mathsf{r} \gets \mathsf{r} -2$
        \EndWhile
        
        \State $\mathsf{lastOrderedRound} \gets \mathsf{v}.\textit{round}$
        \State $\textsc{orderHistory()}$
        
    \EndProcedure
    
    \Statex
    
    \Procedure{\textsc{orderHistory()}}{}
        \While{$\neg \mathsf{orderedAnchorsStack}.\textit{isEmpty}()$} 
        \State $\mathsf{anchor} \gets \mathsf{orderedAnchorsStack}.\textit{pop}()$ 
          \State $\mathsf{verticesToOrder} \gets \{\mathsf{v} \in \bigcup_{\mathsf{r} > 0}
        \mathsf{DAG_i[r]} \mid \textsc{path}(\mathsf{anchor},\mathsf{v}) \wedge \mathsf{v} \not\in $
       $\mathsf{orderedVertices}\}$
          \For{$\textbf{every} ~\mathsf{v} \in \mathsf{verticesToOrder}$ in some deterministic order}
              \State \textbf{order} $\mathsf{v}$
              \State $\mathsf{orderedVertices} \gets \mathsf{orderedVertices} \cup \{\mathsf{v}\}$
          \EndFor
        \EndWhile
        \EndProcedure

\end{algorithmic}
\label{alg:ESBullshark}
\end{algorithm}